\documentstyle[prl,epsf,multicol,aps]{revtex}
\newcommand{\calF}{{\cal F}}
\newcommand{\bfr}{{\bf r}}
\newcommand{\bfu}{{\bf u}}
\newcommand{\rmd}{{\rm d}}
\newcommand{\kB}{k_{\rm B}}
\newcommand{\rmI}{{\rm I}}
\newcommand{\rmN}{{\rm N}}
\newcommand{\rmT}{{\rm T}}
\newcommand{\rmex}{{\rm ex}}
\newcommand{\rmid}{{\rm id}}
\newcommand{\emph}[1]{{\em #1}}
\newcommand{\dir}{./Figs/}
\newcommand{\fig}[4]
{
     \noindent
     \unitlength=1mm
     \begin{picture}(#2,#3)
     \put(10,0){\leavevmode \epsfxsize=#2mm \epsffile{\dir/#1}}
     \end{picture}
   \noindent
#4
}
\begin{document}
\draft
\title{Surface tension of the isotropic-nematic interface} 
\author{Andrew J. McDonald, Michael P. Allen}
\address{H. H. Wills Physics Laboratory, University of Bristol \\
Royal Fort, Tyndall Avenue, Bristol BS8 1TL, U. K.
}
\author{Friederike Schmid\cite{byline}}
\address{Max Planck Institute for Polymer Research \\
Ackermannweg 10, D-55128 Mainz, Germany
}
\date{\today}
\maketitle
\begin{abstract}
We present the first calculations of the pressure tensor profile
in the vicinity of the planar interface between isotropic liquid
and nematic liquid crystal, using Onsager's density functional theory
and computer simulation.
When the liquid crystal director is aligned parallel to the interface,
the situation of lowest free energy,
there is a large tension on the nematic side of the interface
and a small compressive region on the isotropic side.
By contrast, for perpendicular alignment,
the tension is on the isotropic side.
There is excellent agreement between theory and simulation both
in the forms of the pressure tensor profiles,
and the values of the surface tension.
\end{abstract}
\pacs{PACS: 61.30.Cz,68.10.Cr,83.20.Di,83.20.Jp}
\begin{multicols}{2}
The nematic liquid crystal phase is distinguished from the isotropic liquid
by the existence of long-ranged orientational ordering
of the constituent molecules
\cite{degennes.pg:1995.a}.
The structure of the planar interface between coexisting
nematic (N) and isotropic (I) phases is of fundamental interest:
along with the liquid-vapour surface,
it is one of the simplest fluid-fluid interfaces.
Molecular-scale interactions,
capillary-wave fluctuations,
and the coupling between translational and rotational degrees of freedom,
may all influence the interface properties.
The surface excess free energy, 
or interfacial tension $\gamma$,
is a key parameter.
The variation of this quantity with the direction
of molecular ordering in the liquid crystal,
the so-called \emph{director}, 
dictates the molecular alignment which will be preferred by the surface:
the phenomenon of \emph{anchoring},
an essential feature of liquid crystal physics
\cite{jerome.b:1991.a}.
Determining the surface tension 
of the N-I interface for a given molecular model 
by computer simulation,
and predicting it from first principles by theoretical methods,
is clearly desirable.
For a planar interface,
$\gamma$ may be calculated by integrating the
difference between normal $P_\rmN$
and transverse $P_\rmT$ pressure tensor components
with respect to position
\cite{schofield.p:1982.a,rowlinson.js:1982.a}:
$\gamma = \int_{-\infty}^{\infty} \rmd z \left[ P_\rmN(z) - P_\rmT(z) \right]$.

In this paper, we present the first discussion of the
pressure tensor profile through the nematic-isotropic
interface in the framework of a simple density functional
theory in the Onsager approximation
\cite{onsager.l:1949.a},
modelling the molecules as freely-rotating and translating
rod-like rigid bodies.
We shall show that there is a significant difference between the
two situations of most interest,
when the director is oriented respectively
parallel to, and normal to, the interface.
Previous studies
\cite{doi.m:1985.a,chen.zy:1992.a,chen.zy:1993.a,koch.dl:1999.a},
calculating the surface tension from free energy differences,
have indicated that, in the limit of infinite rod elongation,
the minimum surface free energy is obtained when the rods
lie parallel to the N-I interface;
in other words, this interface favours \emph{planar} anchoring.
We provide additional insight into this result,
through an examination of the normal and transverse pressure.

We back up these predictions by carrying out the first computer
simulations to give the pressure tensor profiles for
equilibrium N-I interfaces, confined
between parallel walls under different anchoring conditions,
using the same molecular models as in the theory,
Our results are further supported by 
the first computer simulations of free standing nematic films 
in coexistence with the isotropic phase.
From the simulation viewpoint,
this study is much more challenging than 
determining the surface tension of the liquid-vapour interface
\cite{walton.jprb:1983.a}
since the densities and other properties of the two phases
are very similar.
To our knowledge,
the only other such studies have involved two phases
maintained out of thermodynamic equilibrium
\cite{bates.ma:1997.a,bates.ma:1998.a}.

We employ two very similar molecular models
in this work, in both simulation and theory.
Both models are axially symmetric,
with molecular orientations represented by unit vectors $\bfu_i$
and centre-of-mass positions by vectors $\bfr_i$.
In the first model,
the molecules are taken to be hard ellipsoids of revolution
of axial ratio $\kappa=A/B=15$
where $A$ is the length of the symmetry axis
and $B$ that of the transverse axes;
the bulk phase diagram of this model is well known
\cite{1996.f},
and the interfacial structure has been recently reported by one of us
\cite{2000.d}.
For hard particle models, the temperature enters only trivially into
the thermodynamics, and we define units of energy by setting $\kB T=1$
in this work.  
Length units are fixed by taking $B=1$,
mass units by setting particle mass $M=1$,
and the moment of inertia was fixed at $I/MB^2=50$.
In the second `soft ellipsoid' model,
a continuous pair potential
$v_{12}=4\epsilon(s_{12}^{-12}-s_{12}^{-6}) + 1$,
for $s_{12}<2^{1/6}$, $v_{12}=0$ otherwise,
was used.
Here,
$s_{12}=(r_{12}-\sigma_{12}+B)/B$
is a scaled and shifted separation
where the `diameter'
$\sigma_{12}=\sigma(\hat{\bfr}_{12},\bfu_1,\bfu_2;A,B)$
is a standard approximation
\cite{berne.bj:1972.a,gay.jg:1981.a}
to the contact distance of 
two ellipsoids (again with $A/B=15$ here)
at given relative orientations;
$\bfr_{12}=\bfr_1-\bfr_2$,
$r_{12}=|\bfr_{12}|$ and $\hat{\bfr}_{12}=\bfr_{12}/r_{12}$.
%
The potential is purely repulsive and varies quite strongly 
over a short range of interparticle separation.
We choose an energy parameter $\epsilon=1$,
again take $\kB T=1$, 
and set other parameters as for the hard ellipsoids.

In planar geometry,
with translational invariance in the $x$ and $y$ directions,
we adopt the Irving-Kirkwood convention
\cite{irving.jh:1950.a}
so as to write components of the pressure tensor in the form
\cite{schofield.p:1982.a,walton.jprb:1983.a,walton.jprb:1985.a}
\begin{eqnarray}
\label{eqn:IK}
\lefteqn{P_{\alpha\alpha}(z) = \kB T \rho(z) } \qquad \\
&& - \; \frac{1}{2} \int\rmd\bfr_{12} \int\rmd\bfu_1 \int\rmd\bfu_2
~\left(r_{12}^{\alpha} \frac{\partial v_{12}}{\partial r_{12}^{\alpha}}\right)
\qquad \quad \nonumber \\ & & \times 
\int_0^1 \rmd\zeta ~ 
\varrho^{(2)}(\bfr_{12},z-\zeta z_{12},z+(1-\zeta)z_{12};\bfu_1,\bfu_2)
\nonumber \\ &&
\hspace*{6cm} \alpha = x, y, z \:. \nonumber
\end{eqnarray}
Here, $\rho(z)=\int\rmd\bfu \varrho^{(1)}(z,\bfu)$ is the number density,
$\varrho^{(1)}(z,\bfu)$ the 
(position- and orientation-dependent) single-particle density,
and
$\varrho^{(2)}(\bfr_{12},z_{1},z_{2};\bfu_1,\bfu_2)$ the pair density.
The normal component
$P_\rmN(z) = P_{zz}(z) = P$ 
is independent of $z$ throughout the system,
while the transverse component
$P_\rmT(z) = P_{xx}(z) = P_{yy}(z)$ is not.
Far from the interface, the pressure tensor becomes isotropic,
$P_\rmN(z)=P_\rmT(z)=P$.

For simulations of the soft-ellipsoid system, the
pressure tensor was routinely computed from Eqn~(\ref{eqn:IK})
via the intermolecular forces
\cite{schofield.p:1982.a,walton.jprb:1983.a,walton.jprb:1985.a}.
For the hard-ellipsoid system,
it was calculated by identifying 
nearly-contacting pairs of ellipsoids 
\cite{perram.jw:1984.a,perram.jw:1985.a}.
Such pairs are defined as those which would overlap if their dimensions
are uniformly scaled by a factor $\lambda$, very slightly larger than unity.
Tests were carried out with several choices of $\lambda$:
all of the results here are obtained with $\lambda=1.01$,
the smallest value to give acceptable statistics,
which was also found to be satisfactory in ref~\cite{perram.jw:1984.a}.

The construction and minimization of the Onsager free-energy functional
followed closely the procedure described before
\cite{1999.b,2000.d}.
The excess part of the Helmholtz free energy 
$\calF=\calF^{\rmid}+\calF^{\rmex}$ 
is expressed in terms of
$\varrho^{(1)}(\bfr,\bfu)$:
\begin{eqnarray}
\label{eqn:onsagerF}
\calF^{\rmex}[\varrho^{(1)}]
 &=& - \frac{1}{2}\kB T \int\rmd\bfr_1 \int\rmd\bfu_1 
                      \int\rmd\bfr_2 \int\rmd\bfu_2~
 \\ && \times
               \varrho^{(1)}(\bfr_1,\bfu_1) 
               \varrho^{(1)}(\bfr_2,\bfu_2) 
  ~ f(\bfr_{12},\bfu_1,\bfu_2) \:. \nonumber
\end{eqnarray}
where 
$f(\bfr_{12},\bfu_1,\bfu_2) = \exp(-v_{12}/\kB T) - 1$;
this is the leading term in a low-density virial expansion.
Fluctuation effects, which come in through higher-order terms,
were neglected.  The most important fluctuations are capillary waves,
which only slightly broaden the profiles in small systems,
and do not affect the interfacial tension.
Since $\varrho^{(1)}$ is independent of $x$ and $y$,
$f$ may be integrated over these coordinates.
The logarithm of the
density is expanded in spherical harmonics,
on a discrete grid in $z$
with an interval
$\delta z = B/5 = A/75$: 
\begin{math}
  \ln\varrho^{(1)}(z_i,\bfu) = \sum_{lm} C_{i;lm} Y_{lm}(\bfu).
\end{math}
Periodic boundary conditions apply in the $z$ direction.
In the current application,
this expansion was taken to fourth order in the $Y_{lm}(\bfu)$
(restricting to second order was found to change
the coexistence pressure and densities by $\sim$1\%).
All relevant orientational integrals were expressed
in terms of spherical harmonic coefficients of the density
up to order 10
(reducing to order 8 was found to change
the coexistence pressure and densities by $\sim$0.01\%).
The parameters $C_{i;lm}$ were varied to
minimize the grand potential 
$\beta\Omega[\varrho^{(1)}] \equiv \beta\calF[\varrho^{(1)}]- \beta\mu N$
of an I-N film system
at the coexistence value of the chemical potential $\mu$,
determined by a preliminary bulk calculation.

The normal pressure is constant through the system, and equal to
$-\Omega/V$ in both bulk phases at coexistence.
The transverse pressure tensor profile 
was calculated directly from the minimized
density profiles.  
The definition consistent with Onsager's approximation and the use of the Irving-Kirkwood
convention in the simulations is easily derived from
eqn~(\ref{eqn:IK}),
by inserting the low-density form
$\varrho^{(2)}(\bfr_{12},z_1,z_2;\bfu_1,\bfu_2) \approx
\varrho^{(1)}(z_1,\bfu_1)\varrho^{(1)}(z_2,\bfu_2)\exp(-v_{12}/\kB T)$
and carrying out an integration by parts.
The result is
\begin{eqnarray}
\label{eqn:IKO}
\lefteqn{P_{\rmT}(z) = \kB T \rho(z)} \qquad  \\
&& - \: \frac{1}{2} \kB T
\int\rmd\bfr_{12} \int\rmd\bfu_1 \int\rmd\bfu_2 ~
f(\bfr_{12},\bfu_1,\bfu_2)
\nonumber \\ && \times
\int_0^1 \rmd\zeta ~
\varrho^{(1)}(z-\zeta z_{12},\bfu_1)\varrho^{(1)}(z+(1-\zeta)z_{12},\bfu_2)
\nonumber
\:.
\end{eqnarray}

All simulations were carried out at constant volume. A cuboidal
simulation box was employed in all cases, with periodic
boundaries in the $x$ and $y$ directions.
Conventional, microcanonical ensemble
molecular dynamics (MD) simulations of soft ellipsoids
employed periodic boundaries also in the $z$ direction, 
and used a reduced timestep $\delta t=0.002$.
Several systems of $N=7200$ particles 
in boxes of size $L_x/A=L_y/A=2.5, L_z/A\approx20$
were studied.
At average number densities $0.24 \leq \rho A B^2 \leq 0.255$, 
from an initially isotropic configuration,
one or several nematic films were observed to
assemble spontaneously,
with planar interfaces normal to the $z$-direction
and, in every case,
the director lying parallel to the interface.
In addition,
from an initially nematic configuration aligned in the $z$ direction,
a similar two-phase system nucleated
with the nematic director turned through 90 degrees to lie
parallel to the interface, within roughly 1 million 
time steps. 

One of the systems
at density $\rho A B^2=0.255$,
containing a nematic film of width $\sim 13 B$,
was further equilibrated over 7 million time steps,
monitoring the pressure tensor profiles;
run averages were then accumulated over 4.7 million steps.
In the absence of walls, the interfacial tension can be calculated 
simply from the difference between the system averages 
of $P_\rmN$ and $P_\rmT$. 
This has been done
independently using the transverse component $P_{yy}$, which happens
to be roughly parallel to the director, and the transverse component
$P_{xx}$, which is perpendicular,
giving respectively
$\gamma B^2/k_B T = 0.022 \pm 0.01$, $0.015 \pm 0.01$. 
Thus, the free film simulations yielded clear evidence that the
anchoring at the interface is planar, provided a rough estimate
of the interfacial tension, and
indicated an unexpected difference 
between the transverse stress components $P_{xx}$ and $P_{yy}$.
To provide more convenient control of the director orientation 
relative to the interface,
a more systematic study was therefore conducted in a confined geometry.

Both hard and soft ellipsoids
were simulated using conventional constant-$NVT$ Monte Carlo (MC)
\cite{1989.l}
using $N\approx8000$ particles in
a box of dimensions $L_x/A=L_y/A=2.22$, $L_z/A=22.73$.
Flat walls with a short-range orienting field
at the box ends allowed nematic films of width
approximately $8A$ to be stabilized in a desired director
orientation,
separated by a central isotropic region of width approximately $7A$.
Full details of the system,
together with a description of the film structure,
have been given elsewhere
\cite{2000.d}.
In this paper we focus on the cases of normal (along $z$) 
and in-plane (along $y$)
orientation of the director relative to the interface.
For each model, and each director alignment,
we report the results of averaging over
eight independent runs.
For the hard ellipsoids, more than $10^6$ Monte Carlo sweeps 
were allowed for equilibration,
and the averages were accumulated over a further 
$7.5\times10^5$ sweeps
(one sweep is one attempted move per particle).
For the soft ellipsoids, the corresponding figures are
$3.5\times10^5$ sweeps 
and
$6.25\times10^5$ sweeps.
Profiles of number density $\rho(z)$, 
orientational order parameter $S(z)$
(the largest eigenvalue of the second rank order tensor
\cite{2000.d})
and the pressure tensor
were accumulated.
To facilitate the precise comparison between simulation profiles and
theoretical predictions,
a correction was made for the fluctuations in the interface position,
as described elsewhere
\cite{2000.d}.
Error bars were estimated by assuming a normal distribution
of the results of the eight separate runs.
For the in-plane cases of director alignment along $y$ we detected (as for the free films)
an extremely small discrepancy $P_{zz}=P_{xx}\neq P_{yy}$ in the nematic phase
far from the interface. 
We ascribe this to correlations of aligned
molecules across the periodic box:
to eliminate this finite-size effect from the very sensitive surface tension
calculation we set $P_{\rmT}(z) = P_{xx}(z)$ in these cases.
For normal alignment, we found the $x$ and $y$ directions to be equivalent
and set $P_{\rmT}(z) = (P_{xx}(z)+P_{yy}(z))/2$.

\begin{figure}
\fig{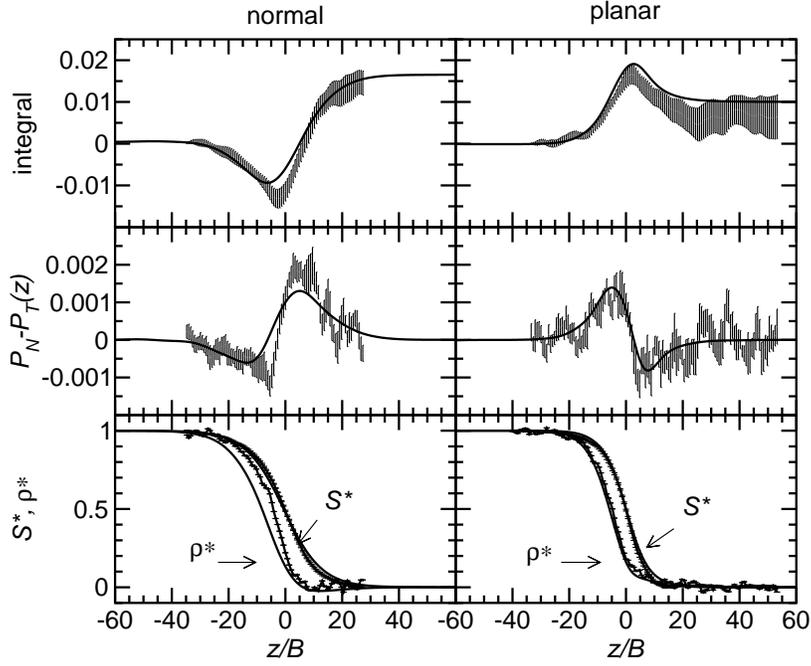}{100}{100}{
\caption[Fig 1]{\label{fig:1} 
Profiles near the N-I interface for hard ellipsoids.
Top: surface tension integral 
$\int_{z_0}^z \rmd z~ P_{\rmN}-P_{\rmT}(z)$.
Middle: pressure tensor difference $P_{\rmN}-P_{\rmT}(z)$.
Bottom: reduced density $\rho^*(z)$ and order parameter $S^*(z)$.
Simulation results are points with error bars,
predictions of Onsager theory represented as lines.
}
}

\end{figure}

\begin{figure}
\fig{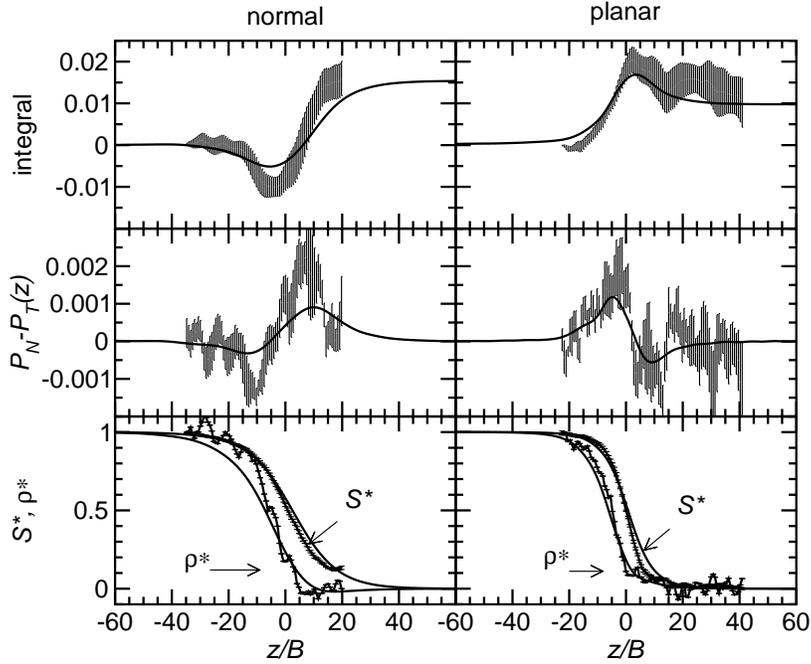}{100}{100}{
\caption[Fig 2]{\label{fig:2} 
Profiles near the N-I interface for soft ellipsoids.
Notation as for Figure 1.
}
}
\end{figure}

Coexistence parameters from theory and simulation are compared
in Table~\ref{tab:coexist}.
As is well known
\cite{1996.f},
for elongation $A/B=15$ the
Onsager theory overestimates
the coexistence pressures and densities by
up to 30\%.
Profiles in the vicinity of the interface are compared
in Figures~\ref{fig:1} and \ref{fig:2}.
To help locate the interface, 
we show the number density curves for both simulation and theory,
scaled by the bulk coexistence values $\rho_{\rmI}, \rho_{\rmN}$ thus:
$\rho^*(z) = (\rho(z)-\rho_{\rmI})/(\rho_{\rmN}-\rho_{\rmI})$,
and similarly for the nematic order parameter $S^*(z)$.
The zero of $z$ is taken at the point of inflection of $S^*(z)$;
as discussed elsewhere
\cite{2000.d},
the density profile lies at $z\approx-A/3$ relative to this
(on the nematic side).
In the figures we show both the pressure tensor difference
$P_{\rmN}-P_{\rmT}(z)$
and the running integral
$\int_{z_0}^z \rmd z~ P_{\rmN}-P_{\rmT}(z)$
taken from a point $z_0$ in the bulk nematic phase far from the interface.
The results, for the two different director alignments,
are dramatically different.
For parallel, in-plane, alignment,
there is a large tension (low transverse pressure)
on the nematic side of the interface, 
followed by a small compressive region (high transverse pressure)
on the isotropic side,
as seen in simulations of the liquid-vapour interface of a simple fluid
\cite{rao.m:1979.a,walton.jprb:1983.a}.
The \emph{surface of tension} 
defined by the first moment of $P_{\rmN}-P_{\rmT}(z)$
\cite{rowlinson.js:1982.a}
lies on the \emph{denser} side of the interface.
For normal alignment,
there is a compressive region on the nematic side, 
and a region under tension on the isotropic side:
the surface of tension lies on the side of the
less dense phase.
Such behaviour is seen in the idealized penetrable sphere model 
of the liquid vapour interface
\cite{rowlinson.js:1982.a}
but to our knowledge this is the first time it has been observed
for a molecular model with realistic interactions.
The step change in the running integrals gives estimates
of the surface tension, reported in Table \ref{tab:coexist}.
We find that the 
planar orientation is favoured, as it has the lower excess free energy;
the same result was obtained by Onsager theory
in the infinite elongation limit
\cite{koch.dl:1999.a},
although the corresponding surface tensions in that case 
(appropriately scaled by $A$ and $B$) are about 40\% larger than here.

Although the simulation results are subject to significant error bars,
we have succeeded in
resolving the variation of transverse pressure through the interface,
and the profiles are very similar, both in form and magnitude,
to those predicted by theory.
Indeed, the level of agreement between theory and simulation results
is remarkable, bearing in mind the
absence of adjustable parameters in the Onsager theory.
In principle, the location of the surface of tension depends on
the convention adopted for the transverse stress,
e.g.\ Irving-Kirkwood vs Harasima \cite{rowlinson.js:1982.a},
but in the current study we find that the profiles are only slightly
dependent on this choice;
moreover, the results for such molecular shapes should be indicative of
the behaviour for more elongated particles,
so we expect them to be rather generally applicable.

This work was supported by 
the Engineering and Physical Sciences Research Council,
the German Science Foundation (DFG),
the Alexander von Humboldt Foundation,
the Leverhulme Trust, 
and 
the British Council.
AJM acknowledges support through grant ORS/99007016 
of the Overseas Research Students Award.
Helpful conversations with
Bob Evans and Ren\'{e} van Roij
are gratefully acknowledged.
Some of this work was performed while MPA and AJM visited
the Max Planck Institute for Polymer Research, Mainz,
and the Institute of Physics, University of Mainz;
the authors are grateful to
Kurt Kremer and Kurt Binder for their hospitality.
%


\end{multicols}
\vfill
\begin{table}
\caption[Tab coexist]{\label{tab:coexist}
Coexistence parameters for hard (HE) and soft (SE) ellipsoids
We report 
coexistence pressure $P_{\rm NI}$,
coexisting number densities $\rho_\rmN$ and $\rho_\rmI$,
nematic order parameter $S_\rmN$
and the surface tensions $\gamma_{\rm NI}$ for 
normal and parallel
director alignment.}
\begin{tabular}{lllllll}
      & $P_{\rm NI} B^3/\kB T$  & $\rho_\rmN B^3$ & $\rho_\rmI B^3$ 
      & $S_\rmN$     & $\gamma_{\rm NI}B^2/\kB T$ (normal) & $\gamma_{\rm NI}B^2/\kB T$ (planar)     \\
HE (Onsager)    & 0.118221 & 0.029645 & 0.025008 & 0.731952 &  0.016           & 0.010           \\
HE (simulation) & 0.099     & 0.0225   & 0.0188   & 0.71     &  $0.014\pm0.003$ & $0.006\pm0.005$ \\
SE (Onsager)    & 0.083528 & 0.022446 & 0.018556 & 0.745852 &  0.015           & 0.010           \\
SE (simulation) & 0.076     & 0.0180   & 0.0149   & 0.74     &  $0.015\pm0.004$ & $0.011\pm0.004$ \\
\end{tabular}
\end{table}
\clearpage


\end{document}